\documentstyle[sprocl]{article}

\begin{document}

\title{\uppercase{Electromagnetic Fields in the QCD Vacuum}}

\author{Johann RAFELSKI}

\address{Department of Physics, University of Arizona\\ Tucson, AZ 85721, USA}

\author{H.-Thomas ELZE}

\address{Instituto de F\'{\i}sica, UFRJ, CP 68528, 21945-970 Rio de Janeiro,
Brasil} 

\maketitle

\abstracts{
Quarks play an active role in shaping the QCD vacuum structure.
Being dual carriers of both `color' and `electric' charges they also respond
to externally applied electromagnetic fields. Thus, in principle, the vacuum
of strong interactions influences higher order QED processes such as
photon-photon scattering.
We survey here the current status of the understanding of the vacuum
structure of strong interactions, and take a fresh look at its
electromagnetic properties.}

\begin{center}
{\it Dedicated to the memory of {\bf Peter A. Carruthers}} \end{center}
\begin{center}
\vskip -11cm
{\bf Presented at the IV Workshop on QCD\\
held at the American University of Paris, France, June 1998\\
 to appear in  Proceedings, H. Fried and B. M\"uller, edts.}
\vskip 10cm
\end{center}

\section{QCD Vacuum}


\subsection{Gluon condensate}


Due to attractive glue-glue interaction inherent in the non-Abelian nature
of color charges, the naive, i.e., non-interacting product wave function of
the vacuum state is known to be unstable\,\cite{MS77}. It is generally
believed that the QCD-originating structures are the source of the
confinement effect which restricts quarks to colorless bound states. 
Many features of the structured vacuum have been studied in past 20 years
with a wealth of methods, but one aspect, the appearance of a  glue
`condensate' field, i.e, vacuum expectation value (VEV) of the gluon
field-correlator in the vacuum state\,\cite{SVZ79} is of particular
relevance in our study. Its value  obtained from QCD sum-rules is
today\,\cite{Nar96} not much different from values first noted nearly 20
years ago\,\cite{SVZ79}:  \begin{equation}\label{Gcon}
\langle V |\frac{\alpha_s}{\pi} G^2| V\rangle
        \simeq (2.3\pm0.3) 10^{-2}{\rm GeV}^4
        =[390\pm12\,{\rm MeV}]^4\,,
\end{equation}
where $\alpha_s=g^2/4\pi$ is the strong interaction (running) coupling
constant, and 
\begin{eqnarray}\label{G2}
\frac12 G^2\equiv\sum_a\frac12 G^a_{\mu\nu}G_a^{\mu\nu}
   =\sum_a[\vec B_a^{\,2}-\vec E_a^{\,2}]\,, \end{eqnarray}
with $a=1, \ldots, N^2-1$ gauge field components for the SU$(N)$ color
charge group. This value  of the glue condensate is in agreement with the
results obtained numerically using lattice gauge theory (LGT)
methods\,\cite{EdGM98}, which in addition provide the shape of the
condensate fluctuations in Euclidean time.

What is the  meaning of the vacuum condensate field? 
The vacuum must be field free, so that the appearance of a field correlator
has no classical analog, it  expresses a Bogoliubov-type 
rotation away from the trivial Fock space state, induced by the
interactions. The effect is often compared to the ferro-magnetism since one
can prove that one of the QCD instabilities is the magnetic gluon spin-spin
interaction\,\cite{Spin}. On the other hand, the confinement effect of color
charged quarks is best understood invoking an anomalous dielectric
property\,\cite{Lee}. 
Both these classical analogs are probably applicable; namely, Lorentz and
gauge invariance property of the vacuum state dictates that the VEV of a
product of two  field operators  satisfies: \begin{equation}\label{EBconG}
\langle  G^a_{\mu\nu}(x) G^b_{\rho\sigma}(x)\rangle=
(g_{\mu\rho}g_{\nu\sigma}-g_{\mu\sigma} g_{\nu\rho})
   \delta^{ab}\langle G^2(x)\rangle /96\,.
\end{equation}
Taking the required contractions and using Eq.\,(\ref{G2}) one finds: 
\begin{equation}\label{EBcon}
\langle \sum_a\vec B_a^{\,2}\rangle=
  -\langle \sum_a\vec E_a^{\,2}\rangle\,.
\end{equation}
We find that, in the vacuum $|V\rangle$, we have $\langle \vec B^{\,
2}_a(x)\rangle$ positive, and $\langle\vec E^{\,2}_a(x)\rangle$ negative.
The signs of the matrix elements arise with reference to the perturbative
non-interacting Fock state, i.e., with respect to products of field
operators which are normal-ordered with respect to the unstructured `free'
state $|0\rangle$, the  so called `perturbative vacuum' 
--- this  use of language is an oxymoron since there can be only one `true'
vacuum $|V\rangle$. 
In the perturbative state $|0\rangle$  the VEV of the gluon field vanishes
by definition because of normal ordering. Without normal ordering there are
(infinite) zero-point fluctuations of the field. 
The interpretation  of Eq.\,(\ref{EBcon}) is that the B-field fluctuates in
the true QCD vacuum $|V\rangle$ with a bigger amplitude than in the
perturbative `vacuum' $|0\rangle$, while the E-field fluctuates with a
smaller amplitude than in the perturbative `vacuum' state. This combination
of effects is a necessary consequence of the symmetries, the primary effect
can be seen as being due to the magnetic gluon spin-spin interaction. There
are obviously many different and equivalent ways to model and understand the
glue condensate,  and we shall not pursue this here in greater detail,
though we shall mention some of the models as needed. 
Our primary interest is in quark fluctuations (condensates as well) to which
we turn our attention now.


\subsection{Chiral symmetry and quark vacuum structure} 


Dirac spinor operators for up and down quarks satisfy the two identities:
\begin{eqnarray}\label{udvec}
\partial_\mu j^\mu_+\equiv \partial_\mu(\bar u\gamma^\mu d)
              =i(m_u-m_d)\bar u d\,,\\
\label{udpseudvec}
\partial_\mu j^{5\mu}_+\equiv \partial_\mu(\bar u\gamma^\mu\gamma_5d)
          =i(m_u+m_d)\bar u \gamma_5 d\,, \end{eqnarray} here $u\,,d$ are
the spinor field operators representing the two light quark flavor fields of
mass $m_u$ and $m_d$ respectively. 
The subscript `+' reminds us that these currents `lift' the `down' quark to
`up' quark, it is an iso-raising operator which increases the electrical
charge by $+|e|$.

We see that when the quark masses are equal the isospin-quark-current
Eq.\,(\ref{udvec}) is conserved, which implies that the Hamiltonian is
symmetric under rotations which mix `u' with  `d' quarks; this is an
expression of the isospin-SU(2) symmetry of strong interactions. 
In case that the quark masses would vanish, by virtue of
Eq.\,(\ref{udpseudvec}) the pseudo-vector isospin-quark-current would be
conserved. This implies that the Hamiltonian would be even more symmetric,
specifically it would be invariant under two chiral symmetries SU(2)$_{\rm
L}\times$SU$(2)_{\rm R}$. 
While $m_{q=u,d}=5$--$15\,{\rm MeV}\simeq 0$ to a good approximation on
hadronic scale of 1 GeV,  there is no sign of the corresponding symmetry,
which would be represented in the hadronic spectrum by doublets of hadronic
parity states. For example, we find only one isospin doublet of nucleons,
not two.
On the other  hand, the  [Adler-Weisberger] sum rules which relate weak and
strong sectors confirm the presence of the intrinsic SU(2)$\times$SU(2)
symmetry in the elementary Hamiltonian. Nambu\,\cite{Nam61} pointed to this
symmetry-breaking in which the ground state breaks the intrinsic (almost)
chiral symmetry of the Hamiltonian.

The way we look at this issue (symmetry breaking by the ground state) today
is as follows:  by virtue of the Goldstone theorem, in the event that the
quark masses vanish exactly, there should be an exactly massless Goldstone
boson with quantum numbers of the broken symmetry, thus spin zero, negative
parity and $I=1$.  However,  since the chiral symmetry of the Hamiltonian
was not exact, the low mass pion state expresses the massless Goldstone
meson of strong interactions. 
In a way one can then see the parity doublets of all strongly interacting
particles as being substituted for by a `direct product' of the Goldstone
boson (pion) with the elementary hadron states. This, in turn, means that
many features of the hadronic spectrum and possibly of the vacuum structure
should strongly depend on the small and seemingly irrelevant quark masses.

A nice illustration of this phenomenon is the observation that in the limit
of vanishing quark masses, the pion mass also vanishes.
We follow the standard approach and  consider the two  matrix elements of
the pseudoscalar and the pseudo-vector between the vacuum state and one pion
state: \begin{eqnarray}\label{pionPV}
\langle \pi^+(p)|\bar u(x)\gamma^\mu\gamma_5 d(x)|V\rangle
   =-i\sqrt{2}p^\mu f_\pi e^{ip_\mu x^\mu}\,,\\ \label{pionPS} \langle
\pi^+(p)|\bar u(x)\gamma_5 d(x)|V\rangle
   =i\sqrt{2}g_\pi e^{ip_\mu x^\mu}\,.
\end{eqnarray}
The right hand side arises by the Lorentz symmetry properties of the (true)
vacuum state $|V\rangle$ and the $\pi^+$-state --- hence $p_\mu
p^\mu=m_\pi^2=139.6$\,MeV. 
The matrix element $f_\pi=93.3$\,MeV is known in magnitude since it governs
the weak interaction decay of pions, and $g_\pi\simeq (350\, \rm{MeV})^2$
follows indirectly from the sum-rules (see below).

Taking the divergence of Eq\,.(\ref{pionPV}) and recalling relation
Eq\,.(\ref{udpseudvec}) we obtain: 
\begin{equation}\label{pion1}
m_\pi^2 f_\pi=(m_u+m_d) g_\pi\,.
\end{equation}
A remarkable consequence of this relation is $(m_u+m_d)\simeq 0.1\, m_\pi$,
an unexpected result, since the current quark mass comes out to be smaller
than the `massless' pion.
Indeed, one finds a quite memorable comment to this point in Weinberg's
treatise~\,\cite{Wei96} (Volume II, p190, bottom) on Quantum Theory of
Fields: `{\ldots  One of the reasons for the rapid acceptance of quantum
chromodynamics in 1973 as the correct theory of strong interactions was that
it explained the  SU(2)$\times$SU(2) symmetry [inherent in Adler-Weisberger
sum rule of 1965] as a simple consequence of the smallness of the $u$ and
$d$ quark masses.}'.

Some of us, who have studied in depth the quark-bag model, will wonder how
the $q\bar q$ structure of the pion, that is so evident in this approach,
can be made compatible with its Goldstone nature. In our opinion the matter
is very simple: there is already a big cancellation of different
contributions which leads in a comprehensive\,\cite{AR82} fit to a pion of
mass ${\cal O}(100)$ MeV. The theoretically, not fully understood, but
critical component in the pion mass is the so called zero-point energy
$E_0\simeq - (1.8\hbar c)/R$, where $R$ is the hadron radius. A small change
in $E_0$ suffices to render pion massless, yet the origin of the sign and
magnitude of $E_0$ remains a theoretical challenge; it can not be understood
as the Casimir energy of the cavity as it has been originally conceived. It
is easy to imagine that $E_0$ expresses aside of the center of momentum
projection correction also the structure of the vacuum, and by virtue of the
Goldstone theorem, it has to be self-consistently-fine-tuned so that for
$m_q\to 0$  the mass of the pion vanishes.

The Nambu-Goldstone structure of the vacuum  was explored intensely in terms
of symmetry relations between current matrix elements (current algebra),
even before QCD was discovered --- for example the Adler-Weisberger sum rule
we mentioned above was a stepping stone to the understanding of the
underlying symmetry. Many rather general vacuum matrix element relations
were obtained, of which, in our context, the most important is the GOR
(Gell-Mann-Oakes-Renner) relation, which adopted to the quark language
reads\,\cite{Ynd83}: \begin{equation}\label{udconGOR}
m_\pi^2f_\pi^2=0.17 {\rm GeV}^4\simeq
-\frac12(m_u+m_d)\langle \bar uu+\bar dd\rangle+\ldots\,.
\end{equation}
Sum-rules have been developed\,\cite{SVZ79} which allow to estimate the
condensate of the Fermi fields from the particle spectra and cross sections,
and the current best value\,\cite{Nar96} is: \begin{equation}\label{udcon}
\frac12\langle \bar uu+\bar dd\rangle\vert_{1\,{\rm GeV}}
  \equiv \frac12\langle \bar q q\rangle\vert_{1\,{\rm GeV}}
   =-[(225\pm9)\,{\rm MeV}]^3\,.
\end{equation}
When combined with Eq.\,(\ref{udconGOR}) one can  estimate the magnitude of
running QCD-current quark masses, which at 1\,GeV scale  are\,\cite{DN98}:
\begin{equation}\label{udmass}
(m_u+m_d)\vert_{1\,{\rm GeV}}\simeq 15\,\mbox{MeV}\,,\qquad 
    m_s\vert_{1\,{\rm GeV}}\simeq 182\,\mbox{MeV}\,.
\end{equation}

\subsection{Quark-gluon relation in the vacuum state} 

We have thus established that the vacuum state has a complex structure in
which significant fluctuations of the quantum gluon and quark fields occur.
Is there a relation between these two different phenomena: i.e.
the chiral structure, and gluon instability? 
One would be tempted to infer that the chiral symmetry-breaking features in
QCD and quark condensation  have little if anything to do with gluon
condensation we described above. However, studies of symmetry restoration at
high temperature\,\cite{lattice}  have yielded contrary evidence: at high
temperature the vacuum structure of QCD melts, as expressed by
Eq.\,(\ref{Gcon}) in terms of the glue condensate, and one reaches the
perturbative vacuum\,\cite{lattice}.
This confinement to deconfinement transformation and the chiral symmetry
restoration, as expressed by the melting of the quark condensate, are seen
exactly at the same temperature.

Furthermore, model calculations\,\cite{DS81,Sot85,EN98}, employing mean
field configurations of gauge fields in the QCD vacuum, invariably suggest
that it is the presence of the glue field condensate which is the driving
force behind  the appearance of the quark condensate. For example, a  model
which employed a self-dual covariantly constant field\,\cite{ES86} for the
non-perturbative gauge field configurations in the structured QCD vacuum
finds that the quark condensation is a minor and stabilizing contributor
(6\%) to the vacuum energy due in its bulk part to the glue degrees of freedom.

The mechanism how this can happen is, in principle, not very difficult to
under\-stand. Schwin\-ger in his seminal paper on gauge invariance and
vacuum fluctuations [see\,\cite{Sch51}, Eq.\,(5.2)] shows already that:
\begin{equation}\label{Psicon}
\langle\bar \psi(x)\psi(x)\rangle\equiv
  \langle\frac12[\bar \psi(x),\psi(x)]\rangle \equiv\Sigma 
       = -{{\partial \Gamma}\over {\partial m}}\,.
\end{equation}
The left-hand side of Eq.\,(\ref{Psicon}) defines more precisely the meaning
of the quark condensate in terms of the  Fermi field operators at equal
space-time points. The right-hand side refers to the effective action
density $\Gamma[A_\mu]$ of fermions in presence of gauge potentials $A_\mu$.

For the case of constant gauge fields, the specific relation Schwinger
obtained for the fermion condensate in the true vacuum, with operators
normal ordered with reference to perturbative vacuum, to first order in
coupling strength but all orders in the gauge field strength is:
\begin{equation}\label{PsiconEB}
-m\frac12\langle[\bar \psi(x),\psi(x)]\rangle^{(1)}=
  \frac{m^2}{4\pi^2}\int_0^\infty\frac {ds}{s^2} {\rm e}^{-m^2s}
  \left[\frac{sE}{\tan sE}\,\frac{sB}{\tanh sB} -1\right]\,.
\end{equation}
where the gauge field configurations are paralleling the Maxwellian electric
$\vec E$ and magnetic $\vec B$ fields: \begin{eqnarray}
B^2&=&\frac{e^2}2\sqrt{(\vec E^{\,2}-\vec B^{\,2})^2
  +4(\vec E\cdot \vec B)^2}- \frac{e^2}2(\vec E^{\,2}-\vec
B^{\,2})\nonumber\\ \label{conB} 
  &\to& |e\vec B|^2\,,\qquad   {\rm for}\ |\vec E|\to 0\,;\\
E^2&=&\frac{e^2}2\sqrt{(\vec E^{\,2}-\vec B^{\,2})^2
  +4(\vec E\cdot \vec B)^2}+ \frac{e^2}2(\vec E^{\,2}-\vec
B^{\,2})\nonumber\\ \label{conE} 
  &\to& |e\vec E|^2\,,\qquad   {\rm for}\ |\vec B|\to 0\,.
\end{eqnarray}
The standard relations of Abelian gauge fields \begin{eqnarray}\label{F2}
\vec E^{\,2}-\vec B^{\,2} = -\frac12 F_{\mu\nu}F^{\mu\nu}
   \equiv -\frac12 F^2\,,\quad
\vec E\cdot \vec B=-\frac14  F_{\mu\nu}\tilde{F}^{\mu\nu}
    \equiv -\frac14 F\tilde{F}\,,
\end{eqnarray}
where:
\begin{equation}\label{Fdef}
F^{\mu\nu}=\partial^\mu A^\nu-\partial^\nu A^\mu\,,\qquad
\tilde{F}^{\mu\nu}= \frac12\epsilon^{\mu\nu\alpha\beta} F_{\alpha\beta}\,.
\end{equation}
apply in suitable generalization for non-Abelian gauge fields. We note that
the effective action is greatly sensitive to details of the gauge field
configuration, and thus its derivative which is the Fermi field condensate,
is probably even more unpredictable as the assumption of the constant (on
scale of the fermion considered, $1/m$) gauge fields is relaxed. Thus the
precise relationship here given is not likely to apply to any realistic
study of QCD structure.

However, the result inherent in Eq.\,(\ref{PsiconEB}) is proving that the
Fermi condensate is driven by the presence of the gauge field fluctuations,
which as we alluded to above, are in turn just little influenced by the
quark condensation. Thus in a self-consistent description of both
condensates the exact solution should be very close to the gauge field
vacuum configuration found without Fermi fields, while quark field vacuum
configuration should be strongly dependent on the presence of the gauge
field fluctuations, which are the key force driving quark condensation.

The Nambu-Goldstone mechanism assures that the quark vacuum must have a
profoundly non-trivial structure which `remembers' even the values of small
quark masses. 
The Nambu-Jona-Lasinio model of strong interactions\,\cite{Nam61,Kle92}
incorporates in  the relevant (spontaneous)
symmetry-breaking features and has  become a frequently studied model of
strong interactions. 
It also offers an opportunity  to explore the interference between
electromagnetic and strong forces, which has been considered with an eye for
the possible chiral symmetry restoration \,\cite{KL89,SMS92,SS97} in fields
of extreme strength.  However, what interests us here is a small distortion
of the vacuum in consequence of a relatively weak Maxwell field being
applied. It remains to be seen if this question, also subtly dependent on
the exact wave function of the vacuum state, can be studied within the
Nambu-Jona-Lasinio approach, or if we need to address these issues within a
new scheme of approach to  QCD.


\section{Electromagnetic Properties of the Vacuum} 


\subsection{Precision vacuum structure  experiments} The presence of the
quark vacuum structure invites naturally an experimental study of the QCD
vacuum employing the electromagnetic quark interactions. When assessing this
option we have to consider that QCD-QED effects are constrained by the well
known QED precision results. 
Consider as example  the QED-vacuum polarization (VP) effect, which on
distance scale  of the electron Compton wave\-length, alters the
$1/r$-nature of Coulomb's  law  by a 0.1\%-relative-strength correction. 
The VP  effect arises as the 
response of the vacuum to  a highly inhomogeneous field-strength, and is
often interpreted as a dielectric particle-hole (electron-positron) photon
polarizibility. Because of its range, the VP-potential can only be detected
in the vicinity of the atomic nucleus. VP-effect does not violate the
superposition principle of electromagnetic fields, thus there is no
possibility for an effective new interaction, an `anomaly' such as
photon-photon scattering, or photon-electromagnetic field interaction. 
VP has been  explored to considerable precision and its agreement with
theoretical expectations speaks for many as evidence against other
measurable electromagnetic vacuum structure.

We need to debunk this myth: the predictability of the usual vacuum
polarization effects in QED arises because  of gauge invariance related to
charge conservation, and the process of charge renormalization. These
symmetry effects combine to `protect' the electromagnetic interactions in
strength and shape. These effects are not present in the  nonlinear higher
order terms which generate otherwise absent interactions, i.e. `anomalies'.
We specifically note the  photon-photon scattering process which is
forbidden in classical electro-magnetism, and which is the key new feature
of QED noted quite early in its development by Euler, Heisenberg,
Kockel\,\cite{HE36} 60 years ago. This effect is not `protected' by
symmetries and can be greatly influenced by the QCD vacuum structure, as we
shall discuss below.

The theoretical question is here in what physical environment we can best
explore such effective interactions. One particularly interesting
environment involves macroscopic Maxwell EM-fields (quasi-constant magnetic
fields and laser fields), fields which  even on atomic scale are extremely
homogenous.  In such situation renormalization absorbs all effects due to VP
and the only effect that remains is the effective higher order interaction,
specifically the effect of light-light scattering.  Much recent effort has
addressed the possibility to study  QCD-QED vacuum structure   using
precision laser-optical QED probes\,\cite{Iac79,Can91,Bak97} in strong
magnetic fields, an approach which promises to test vacuum structure effects
we develop below.

In order to understand the question how quarks could contribute to the
photon-photon scattering in the presence of electromagnetic fields we first
need to remind ourselves how this process works in QED. We than show that
the lowest order Feynman diagrams of QCD in which a virtual up-quark is
immersed in the glue vacuum fluctuations and is polarized by the applied
Maxwell fields, contribute 1000 times the strength of the usual
Euler-Heisenberg term to the light-light scattering process. We then discuss
several different sub-resummation of higher-order Feynman diagrams and
obtain quite different results. This suggests that the non-perturbative
approach, based on partial resummation, has limited validity for
understanding how glue vacuum fluctuations impact the quark fields.

\subsection{EH-QED-effective action: constant Abelian gauge field} 


The effective `one-loop' action $\Gamma^{(1)}$ that is to first order in the
coupling constant $\alpha$ in the  Abelian (QED) theory, and evaluated in
the limit that the (Maxwell) field is constant on the scale of electron's
Compton wave length (which is the situation for all externally applied
macroscopic fields) was studied and its analytical structure and particle
production instability was fully understood in the seminal paper of
Schwinger\,\cite{Sch51}; the non-perturbative aspects of pair production
described in that work provide today basis for particle production dynamics
by gauge fields. 
The form of (renormalized) $\Gamma^{(1)}$ is: \begin{equation}\label{EHfull}
\Gamma^{(1)}_{\rm r}=
 -\frac1{8\pi^2}\int_0^\infty\frac{ds}{s^3}{\rm e}^{-m^2s}
  \left[\frac{sE}{\tan sE}\,\frac{sB}{\tanh sB}
  -1+\frac13(E^2-B^2)s^2\right]\,.
\end{equation}   
We note here two subtractions. The first accounts for the field independent,
zero-point action of the perturbative vacuum, and corresponds to
normal-ordering of the field operators with respect to the no-field
non-interacting vacuum. 
The second subtraction is introduced in order to implement charge
renormalization. 
It assures that for weak fields the  perturbative asymptotic series begins
${\cal O}(E^4,B^4,E^2B^2)$.
This subtraction is accounted for by the subscript `r' in $\Gamma$ in
Eq\,.(\ref{EHfull}).

The analytical structure of the highly non-linear effective action comprises
poles, and the action comprises an imaginary component akin to the situation
one finds for unstable decaying states when interactions are turned on.
Schwinger\,\cite{Sch51} identified the singularities along the real $s$-axis
of the proper-time integral in Eq.\,(\ref{EHfull}) with the pair production
instability of the vacuum, a process in principle possible when potentials
are present that can rise more than $2m$ \,\cite{RFK78}, which is of course
the case in presence of constant, infinite range, electrical fields. This
requirement is consequence of the properties of QED which remains for time
independent fields a stable theory, i.e. there is no spontanous particle
production unless a potential difference (which cannot be gauged away) of
more than $2m$ arises. 
An interesting point in our current discussion is that if we arrange for an
electrical field which is quasi constant on Compton wavelength scale of the
electron, but for which the potential never exceeds this spontanous pair
production  threshold $|V|<2m$, the Schwinger pair production singularities
must vanish, no matter how the potential varies (and fields are time
independent). How this can mathematically occur remains a mystery of higher
order quantum electrodynamic processes, but this observation shows the
sensitivity of the QED to subtle changes of the physical constraints which
seemingly only little impact the mathematical structures.  Similar situation
arises in QCD: Matinyan and Savvidy\,\cite{MS77} identified the magnetic QCD
instability, which  was stabilized by slight inhomogeneity of the vacuum
fields in subsequent work, some mentioned above\,\cite{Spin,Sot85}.

We will need  both the leading and next to leading terms in the (asymptotic)
expansion of Eq.\,(\ref{EHfull}) for small fields: \begin{eqnarray}
\Gamma ^{(1)}_2&=&-\frac{1}{90}\frac{\pi^2}{m^4} [(\frac{\alpha}{\pi}F^2)^2
+\frac74 (\frac{\alpha}{\pi} F\tilde{F})^2]+ \nonumber \\ 
\label{EHtwoterm} 
&+&\frac{1}{315}\frac{\pi^4}{m^8} [4(\frac{\alpha}{\pi}F^2)^3
+\frac{13}2\frac{\alpha}{\pi}F^2(\frac{\alpha}{\pi} F\tilde{F})^2]+\ldots\,.
\end{eqnarray}   
In pure QED, the greatest contribution arises from the smallest mass charged
fermion, the electron. But even for $m_e=0.511$\,MeV, the nonlinearity
arising from electron fluctuations are extremely small compared to the
laboratory fields that can be established. 
The (nearly) observable macroscopic effect of the QED-fermion vacuum
fluctuation is the effect of laser-external field interaction inherent in
the nonlinearity of effective action expansion, Eq.\,(\ref{EHtwoterm}).  
In such experiments\,\cite{Iac79,Can91,Bak97}, the external homogeneous
magnetic  field is crossed by polarized laser light beams. The laser field
corresponds to the visible wavelengths, and thus millions of electron
Compton wavelengths.

One should expect that  this field can be understood as being homogeneous. 
But given the proverbial sensitivity of the effective action, it is not
quite certain that the wave character of the laser beam can be fully
ignored. However, the perturbative expansion which can only have asymptotic
meaning, since the imaginary part of the action is non-analytical for $E\to
0$, promises to be non-sensitive, and thus as long as the perturbative
expansion remains an asymptotic approximation and can be used given the
magnitude of fields used, the consequences of the laser-field scattering
experiment are predictable in the framework of this approximate action.

The EM-field driven fermion condensate is now obtained according to
Eq.\,(\ref{Psicon}):
\begin{equation}\label{PsiconEBs}
-m\langle\bar \psi(x)\psi(x)\rangle^{(1)}= \frac23\, \frac{\alpha}{\pi}F^2
-\frac{1}{90}\frac{\pi^2}{m^4}[4(\frac{\alpha}{\pi}F^2)^2 
    +7(\frac{\alpha}{\pi} F\tilde{F})^2] +\ldots\,.
\end{equation}   
The first term is not dependent on the scale of the Fermi field, as can be
verified on dimensional grounds. For situations where perturbative expansion
is applicable, it provides an interesting relationship between the Fermi and
gauge field condensate, which has been also proposed for the QCD
vacuum\,\cite{DS81}.


\subsection{Magnitude of QCD Vacuum influence on light-light scattering} 

The effective action example derived for non-varying background fields
presented in Eq.\,(\ref{EHfull}) allows to obtain the quark condensate,
Eq.\,(\ref{PsiconEB}),allowing, in principle, to understand how it can be
driven by the  gauge 
field (glue) condensate. In this simple non-varying background field case,
we can fully explore the quark vacuum, including both Abelian  Maxwell (QED)
fields and non-Abelian(strong in comparison) QCD-fields.  We can assume that
the quark structure in the true vacuum is perturbed by Maxwell-EM-fields,
without impacting  significantly other properties of the true vacuum state
which are mostly in the glue sector. In other words, we neglect the
feed-back of the quark condensate polarization by Maxwell fields into the
gauge field QCD vacuum structure.

We now consider how quarks contribute 
to the effective action, remembering to include the quark charge ($q$, in
units of $e$) and quark mass $m_q$, and  we subject quarks also to the glue
condensate along with the relatively small Maxwell field. Up to  factors of
order one, we ought to substitute, in the effective action
Eq.\,(\ref{EHfull}), the pure U(1) Maxwell gauge field by the invariant
combination of U(1) and the condensed SU(3) gauge fields: 
\begin{equation} \label{twofields}
 \frac{\alpha}{\pi}F^2\to 
 \langle\frac{\alpha_s}{\pi} G^2\rangle+ \frac{\alpha}{\pi} q^2F^2\,.
\end{equation}
Here $q=2/3,\,-1/3$ are the fractional quark charges or $q=\pm1$ for
leptons. We are presently rederiving the effective action, incorporating the
$U(1)$ and $SU(3)$ gauge fields from the outset, and consider
Eq.\,(\ref{twofields}) as an {\it ansatz} for now.

As next step, we now adopt the  result Eq.\,(\ref{EHtwoterm}), employing
Eq.\,(\ref{twofields}). We are interested in light-light scattering to
lowest order, thus in the terms comprising four Maxwell fields. 
In addition to the first term in Eq.\,(\ref{EHtwoterm}), we obtain the term
comprising one gluon  condensate component from the second  next to leading
order expansion term given in Eq.\,(\ref{EHtwoterm}) substituting there
Eq.\,(\ref{twofields}). Specifically we obtain: \begin{eqnarray}\nonumber
\Gamma^{(1)}_{4,2}&=&-\frac{1}{90}\frac{\pi^2}{m^4}
[(\frac{\alpha}{\pi}F^2)^2 +\frac74 (\frac{\alpha}{\pi} F\tilde{F})^2]+\\
\phantom{.} \label{EHpertQCD}\\ 
\nonumber &+&\sum_{i=u,d}\frac{1}{315}\frac{q_i^2\pi^4}{m_i^8}
[12(\frac{\alpha}{\pi}F^2)^2\langle \frac{\alpha_s}{\pi} G^2\rangle
+\frac{13}2(\frac{\alpha}{\pi} F\tilde{F})^2 \langle \frac{\alpha_s}{\pi}
G^2\rangle] +\ldots\,.
\end{eqnarray}

The  relevant numerical value determining the relative magnitude of the
usual EH-term and the QCD-vacuum driven term can be now easily obtained
evaluating the ratio of the coefficients of the second and the first term in
Eq.\,(\ref{EHpertQCD}),
\begin{equation}\label{EHQCDfac}
f=\sum_{i=u,d}\frac{24}7\frac{q_i^2\pi^4}{m_i^8}m_e^4
     \langle \frac{\alpha_s}{\pi} G^2\rangle  \simeq {\rm e}^{9\pm2.5}\,,
\end{equation}
a large number indeed in favor of the QCD driven effects, even if the
magnitude of the effects still remains as we believe just below  detection
sensitivity of past experiments! It is interesting to note that the up-quark
dominates the contribution in Eq.\,(\ref{EHpertQCD}) by as much as a factor
1000 over the down quark, since $q_u^2=\frac49=4q_d^2$ and $m_u\simeq
\frac12 m_d\simeq 5\pm1.5$\,MeV.

It is quite rare that the perturbative relative strength is completely
erased by non-perturbative effects, and thus we hope and expect that the
light-light EH-anomalous effective interaction could be indeed governed by
QCD-vacuum fluctuations, rather than QED. On the other hand, the fact that
the up and down quark asymmetry introduces a difference by a factor 1000
suggests considerable opportunity for a change of the magnitude of this
result when the non-perturbative effects are considered.


\subsection{Non-perturbative treatment of glue condensate} \label{NPcg}

The effective expansion parameter of Schwinger action is
$\langle\frac{\alpha_s}{\pi} G^2\rangle/m_u^4$, which is much greater than
unity. Thus we consider the full EH-effective action of quarks in presence
of gluon condensate and  of small Maxwell fields. The problem with this
approach is that the condensate fluctuates on a scale which is 100 times
shorter than the Compton wavelength of the quark, and in fact therefore the
EH-action is quite inappropriate for this limit. 
Yet, let us see what happens 
using  the non-perturbative fermion action $\Gamma^{(1)}$, evaluated
substituting with relation Eq.\,(\ref{twofields}).
For very large (constant) fields, here taken simply as magnetic fields it is
well  known that: 
\begin{equation}\label{Basympt}
\Gamma^{(1)}\to \;\propto B^2\ln B^2 \,.
\end{equation}
Performing the substitution Eq.\,(\ref{twofields}) and expanding through
quadratic terms in $B^2$ we obtain: \begin{equation}\label{EHasympt}
\Gamma^{(1)}\to \;\propto
  {{B^4}\over {\langle \frac{\alpha_s}{\pi} G^2\rangle}}\,.
\end{equation}
We see that the scale of the gluon condensate replaces fermion mass as the
characteristic dimension  in the non-perturbative treatment of the light
quark sector. Thus the perturbative results  hold but with the substitution
$m_q\to 390$\,MeV.

However, this result cannot be correct since, as noted above, it is
mathematically inconsistent, relying on the validity of constant gauge field
approximation, where we know that glue fluctuations are short ranged, and
moreover, we have argued that even subtle deviations from `constancy' can
lead to a qualitative change of the behavior of effective action.  Moreover,
the approach proposed above is ignoring the  physics of the chiral
symmetry-breaking and the Goldstone  vacuum structure. These two crucial
aspects of the QCD vacuum have now been completely lost.

Thus we turn around to see how we can explore the massless (Goldstone) quark
limit within the effective action.  We will only indicate the steps which
are pointing to new and interesting physics. We  follow here the
suggestion\,\cite{SS97} that the main effect of vacuum response to the
Maxwell field is confined to the action of  (massless?) Goldstone bosons
(pions). Then:
\begin{equation}
\Gamma^{\rm QCD}\stackrel{(?)}{=}
    -\frac1{16\pi^2}\int_0^\infty\frac{ds}{s^3}
   {\rm e}^{-m_\pi^2s}\left[\frac{eBs}{\sinh\,eBs}-1\right]\,, \end{equation}
with the Goldstone mass parameter $m_\pi$ comprising all dependences on
vacuum structure. Therefore 
\begin{equation}
\Sigma(B)-\Sigma(B=0)=-\frac{\partial \Gamma^{\rm QCD}}{\partial m_q}=
-\frac{\partial \Gamma^{\rm QCD}}{\partial m_\pi^2} 
   \frac{\partial m_\pi^2}{\partial m_q} =\frac{\partial \Gamma^{\rm
QCD}}{\partial m_\pi^2}
    \frac{\Sigma(B=0,m_q=0)}{f_\pi^2}\,, \end{equation} where we used the
GOR relation\,(\ref{udconGOR}) in the last equality. Moreover using residuum
expansion one easily shows that: \begin{equation}
\lim_{m_q\to 0}\frac{\partial \Gamma^{\rm QCD}}{\partial m_\pi^2}=
\frac1{16\pi^2}\int_0^\infty\frac{ds}{s^2}
   \left[\frac{eBs}{\sinh\,eBs}-1\right]=-eB\frac{\ln \,2}{16\pi^2}\,,
\end{equation}
which leads to the relation:
\begin{equation}\label{SSrel}
\lim_{m_q\to 0}\left[\Sigma(B)-\Sigma(B=0)\right]
=-\frac{eB\ln\,2\,\,\Sigma(B=0,m_q=0)}{16\pi^2f_\pi^2(m_q=0)}\,.
\end{equation}

This result suggests that the mass-scale for the QCD vacuum effect is
provided by $4\pi f_\pi(m_q=0)$. More importantly, it also implies a very
different behavior of the magnetic field dependence of the vacuum action ---
even though relation Eq.\,(\ref{SSrel}) is, strictly speaking, valid in the
Goldstone limit only, which here means that the dominant infrared scale is
the magnetic field and not the mass. An estimate of the two loop corrections
\,\cite{SS97} shows that these contribute $\propto \Sigma_0(eB)^2/(2\pi
f_\pi)^4$ to the quark condensate. One can anticipate a new n-loop series in
powers of $eB/(2\pi f_\pi )^2$. Clearly the full action is in the limit
$m_q\to 0$ completely different from the full EH-from, and one can only
wonder about resummation approaches  based on `constant field' approximations.


\section{Discussion}

We have shown that we cannot, at present, predict the magnitude of the 
QCD-vacuum deformation by an applied Maxwell magnetic field. 
We have pursued several approaches and have seen that the results differ
greatly. A few interesting insights are worth remembering.
The perturbative analysis suggests that the up-quark fluctuations in the
structured QCD vacuum dominate, and dominate the light-light scattering over
the pure QED Euler-Heisenberg effect. We have also seen, assuming that the
glue condensate field is constant on scale of the quark Compton wavelength,
that full re-summation of all possible diagrams including n-loop series is
required for the proper evaluation of the effective action. We also were 
able to confirm\,\cite{SMS92} that non-perturbative but
`constant' condensate assumption implies that the QED-QCD interference
effect is too small to be observed, see section \ref{NPcg}. The assumptions 
entering the derivation  of this result are, however, gravely inconsistent
with e.g. latest QCD-lattice simulations\,\cite{EdGM98}. These  show that
glue condensate fluctuations are occurring on a scale  well below hadronic
size.

This means that the variation of the condensate is an essential and yet
unaccounted factor in non-perturbative evaluation of the QED-QCD effective
action. We also recall that the QCD-glue condensate field is nearly 100
times stronger than the so called critical field strength at which the
perturbative expansion parameter in Eq.\,(\ref{EHfull}) is unity, e.g.
$gE/m^2, gB/m^2$\,, but that it is believed to be a stochastic, fluctuating
vacuum field. Given these observations, we firmly believe that in fact the
physics issue we have addressed remains unresolved and, moreover, one may
completely ignore the popular assumption that the glue condensate provides
the scale in the QCD based light-light scattering effect, and thus renders
it unobservable.

In fact, the first term in lowest order in the glue condensate field, as
here presented, is perhaps the only term that will remain unchanged in the
full treatment of the QCD fluctuations: the perturbative expansion does not
suffer from averaging introduced by the realistic fluctuation properties of
the gauge field condensate, this term is already result of quantum state
averaging. Indeed, one could argue that the fast short-wavelength
fluctuations average out the contributions of the higher order terms to the
action, and thus, in such a case, the  perturbative result we presented
could even turn out to experience only minor corrections. However, extensive
work on stochastic vacuum models \,\cite{Dos87,Sim96} suggest just the
contrary, namely that there are reasons to believe
\begin{equation}\label{factorG}
\langle G^{2n}\rangle =(?) (\langle G^{2}\rangle)^n\,, \end{equation}
which would mean that all higher order terms are significant.
This conjecture is  confirmed by recent LGT results\,\cite{BBV98}.

In summary, the theory of the QED-QCD vacuum structure interference is
not settled at this time.
The perturbative result is most encouraging as point of  departure for more
theoretical studies, which may be done using lattice gauge theory, or
stochastic field methods with QED drift fields. A fully  satisfactory
treatment of this effect requires a totally novel description in which in
our opinion an ultra-strong, but stochastic non-Abelian gauge field with
short correlation length which is driving the effective quark action is
superposed with a weak Maxwell `drift' field. 
We hope to revisit this problem in the near future.

\nopagebreak


\section*{Acknowledgments}

This work was supported in part by the U.S. Department of Energy, grant
DE-FG03-95ER40937, and by CNPq (Brasil), PRONEX-41.96.0886.00, FAPESP-95/4635-0.


%
\end{document}